\pdfoutput=1

\documentclass[aps,prb,reprint,showpacs,superscriptaddress,groupedaddress]{revtex4-1}

\usepackage{graphicx}
\usepackage{amsmath}
\usepackage{amssymb}
\usepackage{dcolumn}
\usepackage{dsfont}
\usepackage{latexsym}
\usepackage{rotating}
\usepackage{bbm}
\usepackage[usenames,dvipsnames]{xcolor}
\usepackage{float}
\usepackage{epsfig}
\usepackage{psfrag}
\usepackage{natbib}
\usepackage{bm}
\usepackage{eucal}
\usepackage{mathrsfs}
\usepackage{braket}
\usepackage{enumerate}
\usepackage{longtable}
\usepackage{subfigure}
\usepackage{bm}
\usepackage{amsfonts}
\usepackage{dcolumn}
\usepackage{bm}
\usepackage{subfigure}
\newcommand{\be}{\begin{equation}}
\newcommand{\ee}{\end{equation}}
\newcommand{\bn}{\begin{eqnarray}}
\newcommand{\en}{\end{eqnarray}}

\usepackage{color} 

\begin{document}

\author{S. Acharya$^{1,2}$}\email{acharya.swagata@phy.iitkgp.ernet.in}
\author{M. S. Laad$^{3,4}$}\email{mslaad@imsc.res.in}
\author{Dibyendu Dey$^{1}$}
\author{T. Maitra$^{5}$}
\author{A. Taraphder$^{1,6}$}\email{arghya@phy.iitkgp.ernet.in}
\title{First-Principles Correlated Approach to the Normal State of Strontium Ruthenate}
\affiliation{$^{1}$Department of Physics, Indian Institute of Technology,
Kharagpur, Kharagpur 721302, India.}
\affiliation{$^{2}$Department of Physics, King's College London, London WC2R 2LS, United Kingdom}
\affiliation{$^{3}$Institute of Mathematical Sciences, Taramani, Chennai 600113, India
}
\affiliation{$^{4}$Max-Planck Inst. fuer Physik Komplexer Systeme, 38 Noethnitzer Strasse,
01187 Dresden, Germany}
\affiliation{$^{5}$Department of Physics, Indian Institute of Technology, Roorkee, Roorkee 247667, India}
\affiliation{$^{6}$Centre for Theoretical Studies, Indian Institute of
Technology Kharagpur, Kharagpur 721302, India.}



\begin{abstract}
The interplay between multiple bands, sizable multi-band electronic correlations and strong spin-orbit coupling may conspire in selecting a rather unusual unconventional pairing symmetry in layered Sr$_{2}$RuO$_{4}$. This mandates a detailed revisit of the normal state and, in particular, the $T$-dependent incoherence-coherence crossover. Using a modern first-principles correlated view, we study this issue in the actual structure of Sr$_{2}$RuO$_{4}$ and present a unified and quantitative
description of a range of unusual physical responses in the normal state. Armed with these, we propose that a new and important element, that of dominant multi-orbital charge fluctuations in a Hund's metal, may be a primary pair glue for unconventional superconductivity. Thereby we establish a connection between the normal state responses and superconductivity in this system. 
\end{abstract}

\maketitle

\section*{Introduction}
Unconventional superconductivity (USC) in layered Sr$_{2}$RuO$_{4}$ has long attracted intense
attention~\cite{maeno} owing to the expectation that it is a superconducting analogue of superfluid
$^{3}$He, with a spin-triplet, odd-parity and chiral order parameter, described by $\Delta({\bf k})\simeq (k_{x}+ik_{y}){\bf z}$~\cite{ivanov}. Such a state would support half-quantum vortices and topologically protected chiral Majorana modes at the sample edges or on domain walls, which would be of fundamental interest. However, a surge of recent data points toward a much more interesting picture. A range of data also attests to the presence of line nodes in the SC gap function~\cite{rice}.  Moreover, presence of sizable spin-orbit coupling (SOC) and multi-orbital character of the system dictate that the multi-band pair function reflect spin-orbital entanglement; $i.e$, it cannot be classified as spin singlet or triplet~\cite{maeno}.  It is fair to say that determination of the actual pair symmetry in Sr$_{2}$RuO$_{4}$ continues to be a fascinating open issue.
Since the USC is an instability of the highly correlated Fermi liquid (FL) state in Sr$_{2}$RuO$_{4}$,
resolution of this puzzle mandates a proper microscopic description of the normal state itself.  Over
the years, extensive experimental studies reveal $(i)$ a $T$-dependent incoherence-coherence (IC-C)
crossover from an incoherent high-$T$ metal to a strongly correlated FL metal.  The latter obtains only
below $T_{FL}\simeq 20-25$~K, while signatures of an unusual metallic state~\cite{maeno} are
evidenced in a range of transport and magnetic fluctuation data for $T>T_{FL}$. $(ii)$ A careful de-Haas van-Alphen (dHvA)~\cite{bergemann} study has mapped out the low-$T(<T_{FL})$ multi-sheeted Fermi surface (FS). Very recent correlated first-principles calculations~\cite{dama1,damascelli1,pavarini} show that it is necessary to adequately include the complex interplay between one-electron band structure, local multi-band
interactions and SOC to get a quantitative accord with dHvA data.  Moreover, the IC-C crossover has
been interpreted in terms of Hund's metal physics~\cite{marvlje1,marvlje2,georges}, where sizable 
influence of Hund
coupling ($J_{H}$) drastically reduces the lattice-FL crossover scale.  However, there is still no
consensus on the details of the normal state (spin and charge) fluctuation spectra that can generate
the specific pair interaction needed to describe the SC pair symmetry constrained by multitude of data.
\section*{Model and Formalism}

   Motivated thereby, we undertake a correlated first-principles theoretical approach using a combination of density functional theory and dynamic mean-field theory (DFT+DMFT) to address these issues.  As an impurity solver, we use continuous-time quantum Monte Carlo (CTQMC)~\cite{ALPS}, extended to low $T$ to access the IC-C crossover.  Armed with excellent accord with the Fermi surface data, we present a detailed quantitative account of the IC-C crossover, and arrive at a very good quantitative accord with a wide range of normal state responses.  We find, for the first time, that a surprising element, namely, nearly
singular inter-orbital charge fluctuations involving the $Ru-4d$ $xy,\, yz,\, zx$ bands, arise due to the complex interplay between orbital-selectivity and SOC. We discuss the implications of this surprising finding for various proposals for USC that emerges in Sr$_{2}$RuO$_{4}$ from the correlated FL at low $T$.

Band structure calculations were performed in the real body-centered tetragonal (BCT-space group Immm/139) structure. We first perform ab-initio density functional theory calculations within GGA and GGA+SO for Sr$_{2}$Ru$O_{4}$ using the full potential linearized augmented plane-wave (FP-LAPW) method as implemented in the WIEN2k code\cite{wien2k}. We perform Wannierization of the Wien2k output bands around the Fermi level via interface programs like WANNIER90~\cite{wan}, WIEN2WANNIER~\cite{wienwan}. This would, in turn, give us the Wannier orbitals around the Fermi level which serve as inputs of the DMFT self-consistency calculation. Three $t_{2g}$ bands, comprising a two-dimensional $xy$-like $\gamma$ band and a quasi-one-dimensional $\beta$ ($xz$-$yz$-like) band, both electronlike, and a quasi-1D holelike $\alpha$-band ($xz$-$yz$-like) cross $E_{F}$.  In line with
several data that point to sizable SOC, the FS topology is correctly reproduced in DFT only when
SOC is included.  However, sizable multi-band electronic correlations are mandatory to obtain
quantitative accord with the dHvA FS~\cite{pavarini}.  The interplay between multiband correlations and SOC will also be crucial to quantitatively describe the IC-C crossover. Thus, use of a realistic  Hubbard model for the $t_{2g}$ bands with real structural input and SOC is mandatory. The model we use~\cite{pavarini} is the three-band Hubbard model:

\be
H=H_{DFT} + H_{loc} - H_{dc}
\ee

\noindent where $H_{DFT}=\sum_{<i,j>}\sum_{\sigma,\sigma'}\sum_{a,b}t_{a\sigma,b\sigma'}^{ij}c_{ia\sigma}^{\dag}c_{jb\sigma'}$,
 $c_{ia\sigma}^{\dag}$ creates an electron in a Wannier state in orbital $a$ with spin-$\sigma$, etc, $H_{dc}$ is
the double-counting correction, and the $t_{a\sigma,b\sigma'}^{ij}$ are the hopping integrals including the SOC term~\cite{pavarini}.  
$H_{loc}$ describes the direct, 
exchange 
, pair-hopping 
and spin-flip 
terms in the onsite $t_{2g}$ basis.  

\noindent The interaction matrix is
\[ \left( \begin{array}{cccccc}
0 & U & U-3J & U-2J & U-3J & U-2J\\
U & 0 & U-2J & U-3J & U-2J & U-3J\\
U-3J & U-2J & 0 & U & U-3J & U-2J\\
U-2J & U-3J & U & 0 & U-2J & U-3J\\
U-3J & U-2J & U-3J & U-2J & 0 & U\\
U-2J & U-3J & U-2J & U-3J & U & 0 \end{array} \right)\] 
in the basis ${xy\uparrow, xy\downarrow, xz\uparrow, xz\downarrow, yz\uparrow, yz\downarrow}$.

\noindent In the $D_{4h}$ site symmetry, the 
$t_{2g}$ states split into a $b_{2g}$ singlet ($xy$) and $e_{g}$ doublet ($xz,\, yz$), with 
$\epsilon_{xz}-\epsilon_{xy}=E_{cf}\simeq 120$~meV being the crystal field splitting. In view of this observation, we work with bare values of $U_{ll',mm'}$ with $l,l',m,m'$=$a,b$ taken to be orbital-independent, with the understanding that these will effectively acquire orbital dependence due to crystal field effect above.

\section*{Results}
We begin by discussing our first-principles correlated approach including SOC, which gives
the actual correlated Fermi surface for Sr$_{2}$RuO$_{4}$. The bare SOC ($\simeq 90-130$~meV) is roughly of the order of the crystal field splitting at the DFT level. Inclusion of sizable $d$-shell correlations has many effects:

\begin{figure}
\includegraphics[width=0.49\textwidth]{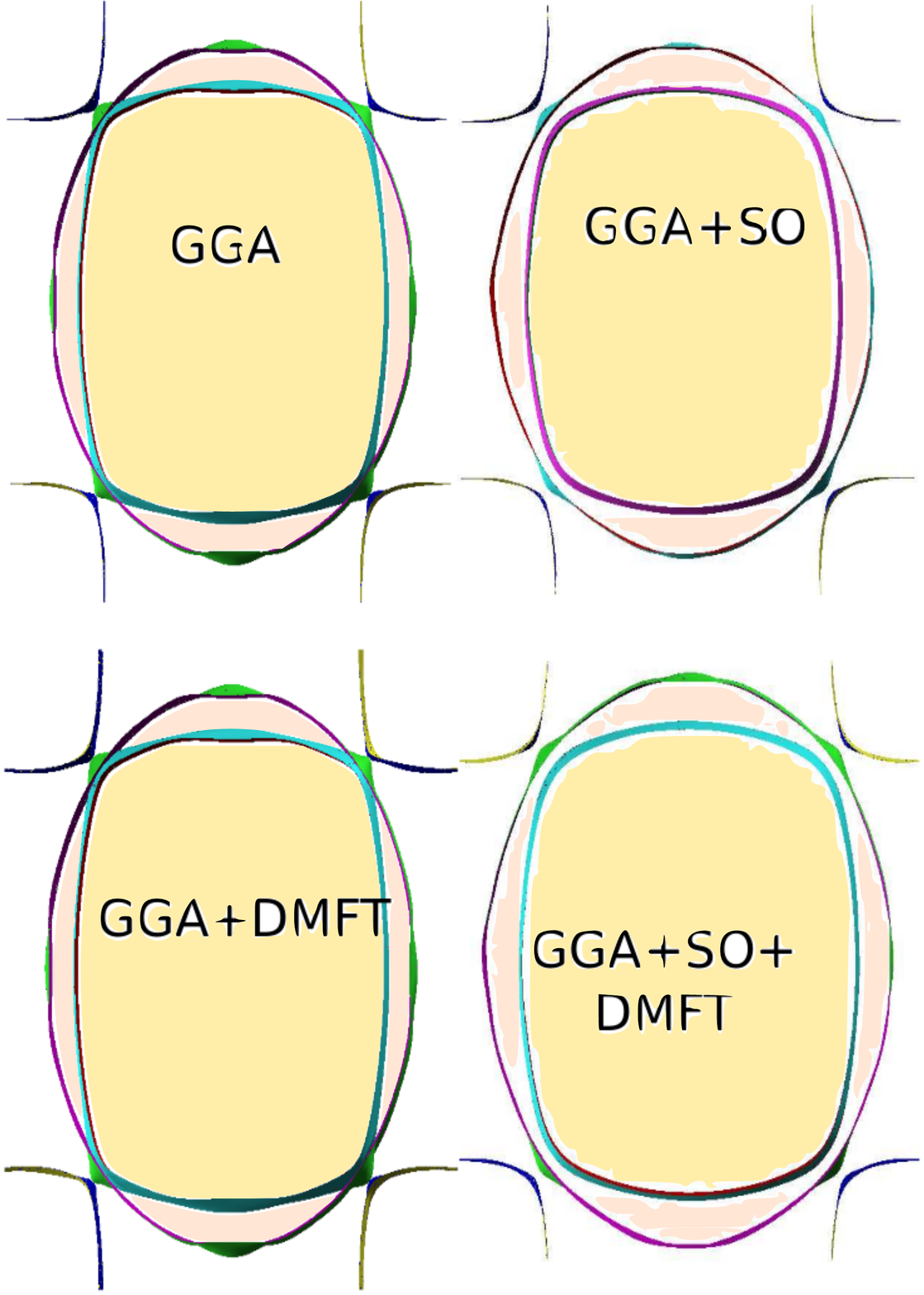}
\caption{Fermi surfaces for Sr$_{2}$RuO$_{4}$ calculated using GGA, GGA+SO, GGA+DMFT, GGA+SO+DMFT. In agreement with Pavarini et al., We find that SOC is essential to derive the quantitatively correct Fermi sheets as seen in dHvA experiment of Bergemann et al.} 
\label{fermi}
\end{figure}

\noindent (i) Correlation effects are larger for the $xy$-orbital states, since they lie lower in energy and are more populated. On the other hand, the smaller band widths of the $d_{xz,yz}$ orbital states mean a larger effective $U/W$ ($W$= bandwidth) ratio for these. This introduces an effective orbital-dependent Hubbard $U$ for the three $t_{2g}$ orbital states, with $U_{xy,xy}>U_{xz(yz),xz(yz)}$, mimicking the situation derived recently by Pavarini {\it et al.} This causes an effective anisotropy in the $d$-shell Coulomb interactions and is indeed the basic mechanism that sets the stage for orbital-selectivity to emerge.

\noindent (ii) Due to inter-orbital charge transfer caused by the interaction terms associated with $\sum_{i,\sigma,\sigma'}n_{i,xy,\sigma}n_{i,xz(yz),\sigma'}$, the bare crystal field splitting is renormalized already at the Hartree level, to an effective value less than its bare DFT value. Since the SOC coupling constant is weakly enhanced by correlations (a simple way to see this is that the static Hartree-Fock contribution from the Hund term directly renormalizes the bare SOC, so $\lambda_{soc}^{eff}=\lambda_{bare}+J\langle c_{ia\sigma}^{\dag}c_{ib\sigma'}\rangle$), the ratio of the effective SOC to the effective crystal field splitting is enhanced upon switching on electronic correlations.  This means that it is no longer possible to disentangle orbital and spin degrees of freedom, a feature which must have far-reaching consequences for the detailed symmetry of the pair wavefunction in Sr$_{2}$RuO$_{4}$.

Both these effects directly bear upon the renormalized electronic structure and the Fermi surface topology.  In Fig.~\ref{fermi}, we exhibit the GGA, GGA+DMFT, GGA+SOC and GGA+SOC+DMFT Fermi surfaces.
It is clear that GGA+DMFT alone gives Fermi surfaces in discord with dHvA data, and that
inclusion of SOC is mandatory to obtain correct Fermi surface sheets with regard to their
shapes and size~\cite{pavarini}.

In addition, these changes in the bare DFT parameters cause changes in
dynamical spectral weight transfer, caused by higher order terms in the self-energy in DMFT.
Thus, effective mass enhancements will be orbital-dependent in the low-$T$ FL state,
as is known~\cite{bergemann}, and different bands are narrowed down to differing extent. The average effective mass
enhancement at $30$~K is $O(3-4)$, in excellent accord with both dHvA estimates and specific heat data.
The coherent part of the ``renormalized band structure'' could be tested against ARPES band structures in future.
More importantly, we find that the orbital character of the three $t_{2g}$ bands is sizably ${\bf k}$-dependent
due to momentum-dependent spin-orbital entanglement.  This is a crucial input when one considers the construction of a ``pair interaction'' to obtain USC: the pair interaction must also reflect this spin-orbital entanglement.  We leave this aspect for future studies.

\begin{figure}
\includegraphics[width=0.49\textwidth]{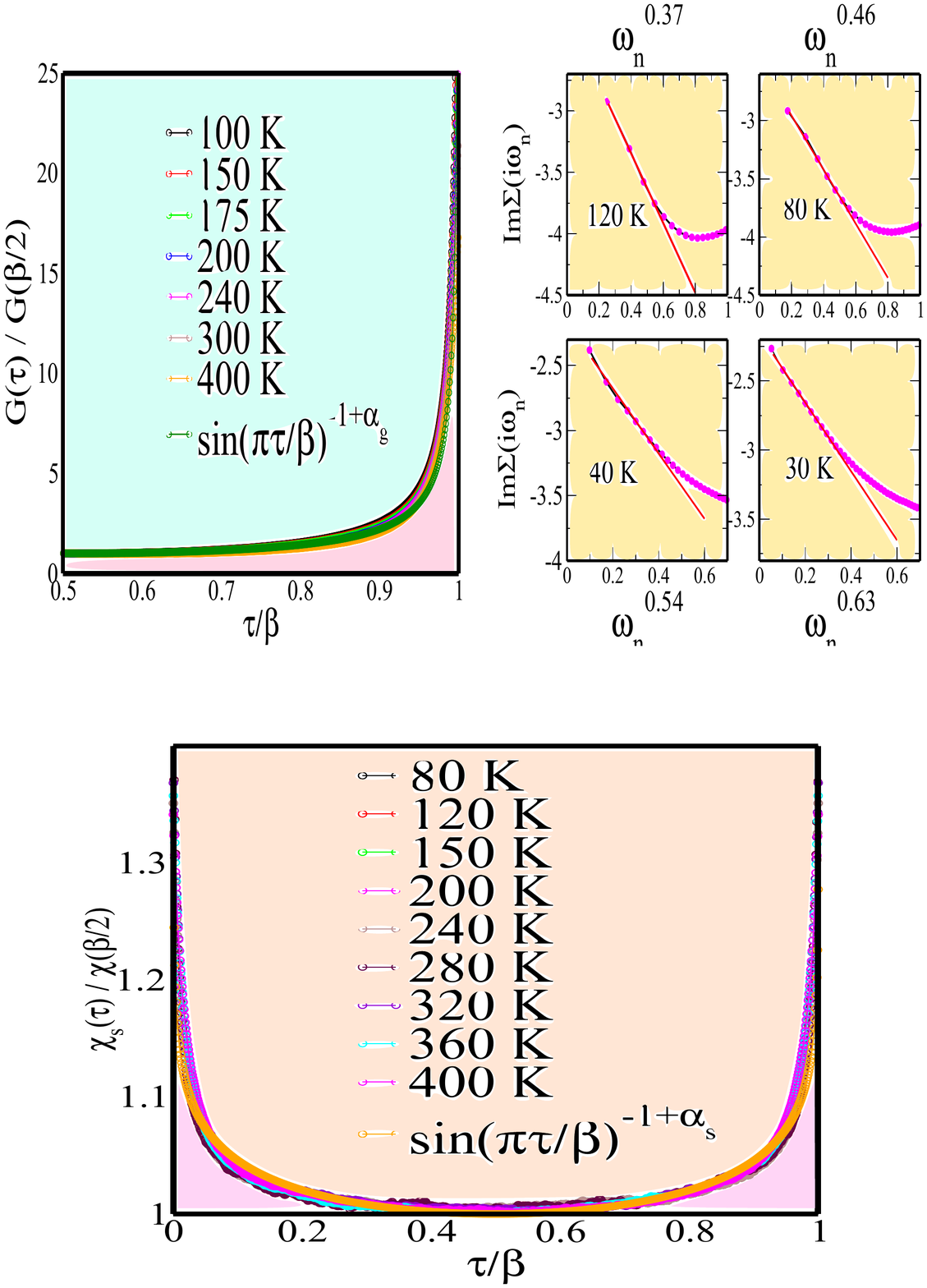}
\caption{Imaginary parts of the scaled one-electron Green function (top left) and the scaled dynamical spin susceptibility (bottom) as functions of $\tau/\beta$.  For $T>T_{FL}\simeq 30$~K, both exhibit clear local quantum-critical scaling behavior with different exponents (see text), implying strange metallicity in Sr$_{2}$RuO$_{4}$ above $30$~K.  The imaginary parts of the self-energy (top right) as a function of $i\omega_{n}$: clear anomalous power-law behavior up to high energy $O(0.5)$~eV further testifies to anomalous metallicity for $T>30$~K.}
\label{scaling}
\end{figure}

   Armed with this accord, we now study the normal state responses in Sr$_{2}$RuO$_{4}$ in detail.  In Fig~\ref{scaling}, we show the full one-electron local Green function in imaginary time, $G(\tau)$ as a function of
$T(>T_{FL})$.  Very good collapse of all curves on to a single curve is clear and, upon careful fitting with a ``local'' quantum-critical form, $G(\tau)/G(\beta/2)\simeq c_{0}$sin$(\pi\tau/\beta)^{-(1-\alpha_{g})}$ with $\alpha_{g}=0.32$.  This conformal-invariant form is that expected for a locally quantum-critical metal, where the infra-red pole structure of $G(\omega)$ in a FL metal is supplanted by an infra-red branch-cut continuum form with anomalous fractional exponents.  Correspondingly, Fig~\ref{scaling} shows that the self-energies also exhibit anomalous behavior:
Im$\Sigma(\i\omega)\simeq (i\omega)^{1-\eta(T)}$ up to rather high energies.  Remarkably, the dynamical spin
susceptibility also exhibits similar scaling behavior:  in Fig.~\ref{scaling}, clear collapse of $\chi_{s}(\tau)$ to a universal scaling function, given by $\chi_{s}(\tau)/\chi_{s}(\beta/2)\simeq c_{1}$sin$(\pi\tau/\beta)^{-(1-\alpha_{s})}$, with $\alpha_{s}\simeq 0.93$ is seen.  This is precisely the form expected for an intermediate $T(>T_{FL}$) state exhibiting conformal invariance.  Using a well-known trick, the real-frequency local dynamical spin susceptibility is then obtained as Im$\chi_{s}(\omega,T)\simeq c_{1}T^{\alpha_{s}}F(\omega/T)$, where $F(y)$ is a universal scaling function of $y$~\cite{si}.

   Microscopically, a reason for these findings is that the Hubbard $U=2.3$~eV is larger than the 
band-width of xz,yz bands ($W_{xz,yz}=1.4$~eV), while it is comparable to that of the $xy$-band.  
Thus, depending upon the orbital, one is effectively either in the intermediate 
($xy$) or the strong-coupling ($xz,yz$) of the three-band Hubbard model, with non-integer occupation of each orbital.  
In the $xz,yz$-sector, we now expect a tendency to have orbital-selective Mott-like states.  
At high $T$, the inter-orbital hybridization is effectively rendered irrelevant by sizable $J_{H}$ (Hund's metal scenario), 
and thus one deals with an effective situation where there is strong scattering between metallic ($xy$) and 
effectively localized ($xz,yz$) carriers.  In the local impurity model of DMFT, 
this implies that the latter cannot recoil during an interband scattering process, 
leading to emergence of recoil-less X-ray edge physics.  
The high-$T$ infra-red singularities in one- and two-particle propagators we find in DMFT are thus associated with local processes akin to those occurring in the seminal orthogonality catastrophe~\cite{anderson}.  

   Remarkably, these results offer direct insight into the high-$T$ anomalies 
characteristic of $Sr_{2}RuO_{4}$.  First, transport can now be rationalized solely 
in terms of the structure of $G_{loc}(\omega)$ or
 $\Sigma_{loc}(\omega)$, since local irreducible vertex corrections are negligible~\cite{biermann} in multi-orbital DFT+DMFT.  
Explicitly, using results above, we find that the $dc$ resistivity, 
$\rho_{dc}$(T)$\simeq T^{2(1-\alpha_{g})}$
$\simeq T^{1.0}-T^{1.5}$, which is qualitatively consistent with observations above $T_{FL}\simeq 30$~K~\cite{maeno} right up to $900$~K.  Correspondingly, the optical conductivity,
 $\sigma(\omega)\simeq \omega^{-2(1-\alpha_{g})}\simeq \omega^{-1.0},\omega^{-1.3}$.
This also implies anomalous energy-dependent scattering rate, $\tau^{-1}(\omega)\simeq \omega^{1.0}, \omega^{1.3}$, and energy-dependent effective mass enhancement, $m^{*}(\omega)/m_{b}\simeq \omega^{-1.0}, \omega^{-1.3}$.  Remarkably, both these features seem to be in accord with data (see Fig.(4) of Katsufuji {\it et al.}~\cite{katsufuji}, and our results suggest a re-interpretation of the $T>T_{FL}$ transport data within an
incoherent metal scenario (indeed, no FL contribution is seen down to $0.03$~eV in
optical data for $T>T_{FL}$).  Moreover, the local critical form of the dynamical spin
susceptibility permits rationalization of the anomalous neutron and NMR results
above $T_{FL}$ as follows: $y=\omega/T$-scaling in $\chi_{s}(\omega,T)$ with
an exponent $(1-\alpha_{s})\simeq 0.93$ implies that $\omega^{0.933}$Im$\chi_{s}(\omega,T)$ must be a
universal scaling function of $y$ for $T>T_{FL}$.  This is fully borne out by neutron scattering
data above $30$~K~\cite{braden}, which is precisely the crossover scale for FL behavior.
In addition, $\alpha_{s}=0.933$ is also close to the exponent of $1.0$ used to fit the neutron scattering
intensity for $\omega > 2.0$~meV~\cite{sidis,werner}.
Finally, the full-width at half-maximum, related to the damping of magnetic excitations
must scale as $(\omega,T)^{1-\alpha_{s}}\simeq (\omega,T)^{0.933}$, again in excellent accord with data.
The NMR spin relaxation rate should vary as $1/T_{1}=T\lim_{\omega\rightarrow 0}\frac{Im\chi_{s}(\omega,T)}{\omega}\simeq T^{-0.93}$,
implying a sizably $T$-dependent (but increasing with reduction of $T$) $1/T_{1}$ for $T>T_{FL}$: this is also
not inconsistent with data~\cite{ishida} for ${\bf H} ||c$.  Thus, good accord with a range of normal state responses lends strong support to an intermediate-$T$ local non-FL ``strange'' metallic state in Sr$_{2}$RuO$_{4}$.

   In contrast to high-T$_{c}$ cuprates however, $Sr_{2}RuO_{4}$ shows a
smooth crossover from the high-$T$ non-FL metal to a correlated FL metal below $T\simeq 30$~K.
Understanding the nature of this crossover is mandatory to uncover the microscopics of
USC setting in below $T_{c}=1.5$~K.  We have done a careful extension of
CTQMC to reach $T\simeq O(12)$~K, allowing us to study this IC-C crossover in detail.
In Fig.~\ref{scaling}, we exhibit evidence for gradual restoration of FL-like metallicity at
low $T$, which is signalled by the fact that the $\omega=0$ intercept in Im$\Sigma(i\omega)$
approaches zero as $T$ is lowered.

This is also borne out by the observation that the 
orbital-dependent effective masses acquire sizable enhancement (SI2), 
become almost $T$-independent, along with drastically reduced scattering rate 
below $T\simeq 23$~K, in excellent accord with indications for a low-$T$ correlated FL metal 
from optical data~\cite{katsufuji}.  
In fact, the effective masses for all orbitals are enhanced by a factor of about $3-4$ at low $T$, 
in nice accord with specific heat data~\cite{werner}.  Since the out-of-plane 
resistivity also acquires a $T^{2}$ dependence below $T_{FL}$, this evidences that a dimensional crossover is implicated in the non-FL-to-FL crossover.  In a quasi-2D correlated FL, this crossover should occur when $k_{B}T\simeq t_{\perp}$, the one-electron hopping between weakly coupled layers.  One might then wonder whether (and how) the strong incoherent metal signatures found for $T>T_{FL}$ influence this crossover.  We now provide a detailed analysis of this aspect, showing how ``high-$T$'' incoherence indeed dramatically affects details of the low $T$ FL state in ways incompatible with a simple one-electron band-structural view.  

  Physically, at low $T$, the interband hybridization eventually gets relevant at a scale pushed dramatically downward by Hund metallicity, introducing recoil into the local impurity model, and cutting off the infra-red singularities found for $T>T_{FL}$.  To consider this crossover as a function of $T$, we
observe that: $(i)$ the interlayer hopping, $t_{\perp}\simeq 0.02eV (200K)$<< $t_{a\sigma,b\sigma'}$~\cite{maeno}, even in the renormalized DMFT electronic band structure, and $(ii)$ at high $T$, the $c$-axis resistivity shows insulator-like behavior in contrast to the bad-metallic in-plane resistivity.  These arguments
show that a dimensional crossover from effectively decoupled $2D$ layers (at high $T$) to an anisotropic $3D$ state is implicated in the IC-C crossover.  
A description of the effect of $t_{\perp}$ requires consideration of a model
with coupled $RuO_{2}$ layers.  Since the interlayer hopping for the
$d_{xy}$ band is much smaller than for the $d_{yz,zx}$
bands in the undistorted BCT structure, we are led to consider the effective
 model of two coupled quasi-1D xz,yz bands tom facilitate the description of the IC-C crossover as a dimensional crossover, driven by increasing relevance of coherent one-electron inter-orbital mixing at lower
$T$:

\be
H=\sum_{i,a\neq b=xy,yz,zx}H_{1D}^{a} - \sum_{i,a,b,\sigma}t_{\perp}(C_{ia\sigma}^{\dag}C_{ib\sigma}+h.c)
\ee
where $\nu,\nu'=xz,yz$.  We now discuss the IC-C crossover and the associated dimensional crossover with a somewhat subtle analytic framework.  For a system of coupled $1D$-like chains as above, a description of this crossover by
perturbation theory in $t_{\perp}$ in the non-FL metal is beset with difficulties, and is valid only in
the non-FL regime, but fails to reproduce the FL regime.  Perturbation approaches in
 interaction, beginning from the free band structure work in the FL regime, but fail in the non-FL
regime.  An attractive way out is provided by a close generalization of a non-trivial argument developed in the
context of coupled Luttinger chains~\cite{biermann}.  In our case, each RuO$_{2}$ layer is connected via
$t_{\perp}$ to $z_{\perp}$ nearest neighbors, with $z_{\perp}\rightarrow\infty$.

Rigorously, one needs a numerical solution for the full local propagator, $G(k,\omega)$, which we take from our DFT+DMFT calculation.  Using this input, we can draw qualitative conclusions regarding the effect of non-FL metallicity at high-$T$ on the non-FL to FL crossover at lower $T$ as follows.  In the non-FL regime, for each of the quasi-1D xz,yz bands, the in-plane self-energy 
$\Sigma_{\nu}(\omega)\simeq (t(\omega)/t)^{1/(1-\alpha)}$~\cite{biermann}.  It is clear that $t_{\perp}$ becomes relevant, inducing the dimensional crossover when $t_{\perp}^{a,b}>\Sigma_{a}$, yielding the crossover scale $E^{*} \simeq t_{\perp}(t_{\perp}/t)^{\alpha/(1-\alpha)}$.  With $\alpha=0.32$ in our case, this yields $E^{*}\simeq 40$~K, in good accord with the numerical estimate of $23$~K in numerics .
   
  Once $T<E^{*}$, one ends up with an anisotropic correlated FL metal.  In particular, when $t_{\perp}<<t$ and at low energies, all one-particle quantities obey the scaling $\omega'=\omega/E^{*}$, and $T'=T/E^{*}$; i.e.,
$t\Sigma(\omega,T)=E^{*}t_{\perp}\Sigma'(\omega',T')$ and $tG(\omega,T)=(E^{*}/t_{\perp})G'(\omega',T')$ where $\Sigma$ and $G$ are universal
functions associated with the crossover.  A low-frequency expansion of $\Sigma$
in the FL regime gives the quasiparticle residue $Z \simeq (t_{\perp}/t)^{\alpha/(1-\alpha)}=E^{*}/t_{\perp}$.
The inter-layer resistivity, $\rho_{\perp}(T)/\rho_{0}=(t/E^{*})R(T/E^{*})$ with
 $R(x<<1) \propto x^{2}$ and $R(x>>1) \propto x^{1-2\alpha}$.
And the resistivity enhancement,
$\rho_{\perp}(T)/ \rho_{0} =A (T/t)^{2}$ with $A=(t/t_{\perp})^{3/(1-\alpha)}$.
The resulting anisotropy of the Woods-Saxon ratio,
$A_{c}/A_{ab}=(a/c)^{2}A \simeq 1000$
for $\alpha=0.32$, which is indeed in the right range~\cite{maeno}.
Finally, the $c$-axis optical response is incoherent above $E^{*}$,
with a coherent feature carrying a relative weight $\simeq Z^{2}$ appearing at
low-$T$, again in qualitative agreement with observations~\cite{katsufuji} .
An obvious inference from the above is that increasing $T$ should lead to a
disappearance of the quasicoherent features in photoemission.  This may already have been observed experimentally~\cite{damascelli2}.  

  The above arguments show how high $T$ incoherence stemming from effectively $1D$ xz,yz states drastically affects details of the IC-C crossover.  At low-$T$, in the correlated FL regime, DMFT(CTQMC) provides a
consistent description of electronic correlations and lead to very good quantitative accord with dHvA and mass enhancements, as shown above.  In our
picture, therefore, the effective mass enhancement arises from renormalization
effects caused by local multiband electronic correlations (DMFT).  This completes our qualitative description of the non-FL to FL crossover, where the ``high''-$T$ non-FL metal is taken to be the local critical metal.  Thus, the physical mechanism for this crossover (which is also a dimensional crossover) is the increasing relevance of the inter-layer one-electron hopping at lower $T$, since the SOC seems to become relevant at lower $T$ as reflected in the in-plane versus out-of-plane spin susceptibility anisotropy~\cite{maeno}.  
\begin{figure}
\includegraphics[width=0.49\textwidth]{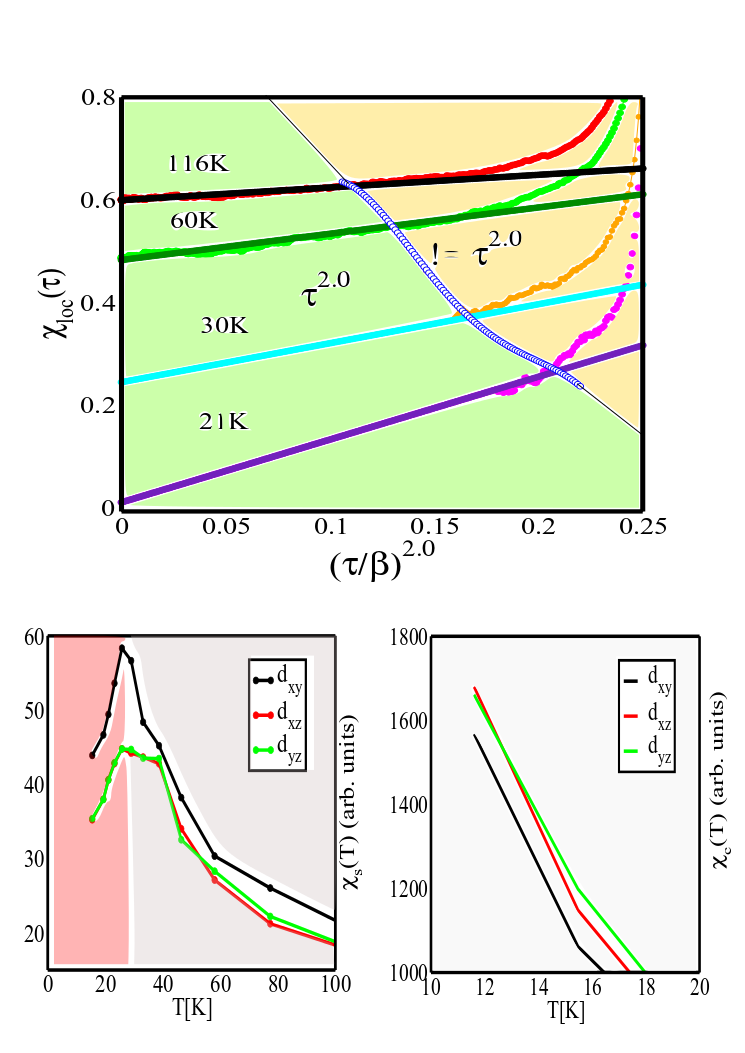}
\caption{Gradual restoration of correlated Landau Fermi liquidity as $T$ is lowered below $30$~K, as seen by the fact that the intercept on the $\chi_{loc}(\tau)$ axis vanishes at $\simeq 21$~K (top). The $\chi_{loc}(\tau)$ is $\tau^2$ (Fermi Liquid like) for finite low energies and finally deviate from $\tau^2$ across the blue boundary where incoherence sets in. With lower temperatures the FL liquid behavior sustains over larger energy ranges, showing predominance of coherence. The incoherence-coherence crossover is reflected in the change in the $T$-dependent band-resolved spin susceptibilities around $30$~K (bottom left). We show that the orbital-resolved charge susceptibilities are enhanced relative to the spin susceptibility at low $T$, making them attractive candidates for an electronic pair glue (bottom right).}
\label{crossovers}
\end{figure}

   To further characterize this crossover and search for a dominant electronic pair ``glue'', we computed the 
$T$-dependent local spin ($chi_{s}(T)$) and orbital-resolved charge ($\chi_{a,ch}(T)$, with $a=xy,yz,xz$) susceptibilities.  In Fig.~\ref{crossovers}, we show $\chi_{s}(T)$ from high- up to low $T$.
Clear Curie-Weiss like behavior at high $T$ smoothly crossing over to a high but relatively $T$-independent
value at low $T<23$~K is a manifestation of the non-FL-to-FL crossover.  Moreover, we find a high-spin state ($S=1$)
on Ru, and thus a Hund metal, arising from sizable Hund coupling.
Several points are in order: $(i)$ details of the $T$-dependence of $\chi_{s}(T)$ are in good
accord with data if one assumes~\cite{sidis} a ``relaxor'' form $\chi_{s}(q,\omega)=f(q)\chi_{s}^{loc}(\omega)$
with $\chi_{s}^{loc}(\omega,T)$ evaluated within DMFT as above.
Due to enhancements at incommensurate ${\bf q}_{in}$ coming from the quasi-1D xz,yz bands
(which enter via the factor $f(q)$ in an RPA like view), the relevant quantity to
compare is $\chi_{s}({\bf q}_{in},\omega,T)$.  Assuming, in spirit of DMFT,
that the $T$-dependence is dominantly in $\chi_{s}^{loc}$, we find surprisingly good
accord with data over the whole $T$ range, right down to $23$~K, where $\chi_{s}(q_{in},T)$ flattens out.
However, the orbital-resolved charge susceptibilities spring a
surprise: in Fig.~\ref{crossovers}, we show that $\chi_{a,ch}(T)$ for $a=xz,yz$ is
enhanced relative to $\chi_{xy,ch}(T)$, and, more importantly, that $\chi_{a,ch}(T)$ always increases at lower
$T$ without flattening out, at least down to $12$~K.
We expect $\chi_{a,ch}$ to eventually approach an enhanced $T$-independent value at low $T$ (since the metallic state
 immediately above $T_{c}=1.5$~K is a FL), but we are unable to access this possibility numerically below $12$~K.
It is also noteworthy that $\chi_{xy,ch}$ is comparably enhanced at low $T$.  Our finding implies that interband
charge fluctuations arising from the quasi-1D $\alpha,\beta$ FS sheets (xz,yz orbital character) are the ``softest''
(the incommensurate spin susceptibility ``flattens'' out to sizably lower values at low $T$), while the charge
fluctuations in the xy-band closely track them, strongly suggesting a novel possibility for the SC mechanism, were these to be implicated in the SC pair glue.  It is crucial to note that clear orbital ``selectivity'' reflected in our results is a consequence of the dynamical interplay between (DFT) structure (in particular, the crystal field splitting between the xz,yz and xy orbitals in the $D_{4h}$ structure), sizable multi-orbital correlations and spin-orbital entanglement due to SOC enhanced by correlations.  Thus, our analysis bares a new, hitherto unappreciated factor: the ``normal'' state above $T_{c}$ maybe close to being electronically ``soft'' in the inter-orbital charge channel.  A wide range of normal state responses we describe above lends strong credence to the feasibility of this conclusion.

We now qualitatively detail how these findings have novel implications for mechanism(s) for USC in 
Sr$_{2}$RuO$_{4}$.  The nearly soft interband charge fluctuation modes found above must overwhelm the
incommensurate spin fluctuations at ,low $T$ (just above $T_{c}$), raising the novel, hitherto neglected
possibility that these might be primary sources for the ``pair glue'' leading to USC.
If this be the case, our results provide support to mechanism(s)~\cite{raghu,simon,ronny}
Further, strong SOC also implies that these multi-orbital charge fluctuations are also intrinsically entangled
 ~\cite{damascelli1} with mixed spin singlet-triplet magnetic fluctuations.  While Raghu et al. claim
that USC primarily arises from the quasi-1D xz,yz bands, and that the interband proximity effect
induces a secondary SC gap in the $xy$ band~\cite{agterber}, Simon et al. and Thomale {\it et al} claim, 
within a weak-coupling RG procedure, that all three $d$-bands are implicated in SC.  However, our results,
along with earlier local density approximation (LDA)+DMFT ones~\cite{pavarini,marvlje2} show that Sr$_{2}$RuO$_{4}$ is more aptly characterized as a sizably-to-strongly correlated system.  The RG procedures hitherto used for study of pair
symmetry are valid in the weak coupling ($U<<W$) limit, but would be inadequate when $U\simeq W$, as appropriate here.  Strictly speaking, it is possible that the weak coupling calculations identify the 
correct pair-symmetry if one could invoke analytic continuity from weak to the intermediate-coupling
state: this maybe possible for Sr$_{2}$RuO$_{4}$, since the low-$T$ state immediately above $T_{c}$ is a correlated FL, which is analytically continuable in the Landau sense from a weakly correlated metal.
Such a study, however, remains to be done, especially within first-principles correlated approaches, and is out of scope of our present work.
\begin{figure}
\includegraphics[width=0.49\textwidth]{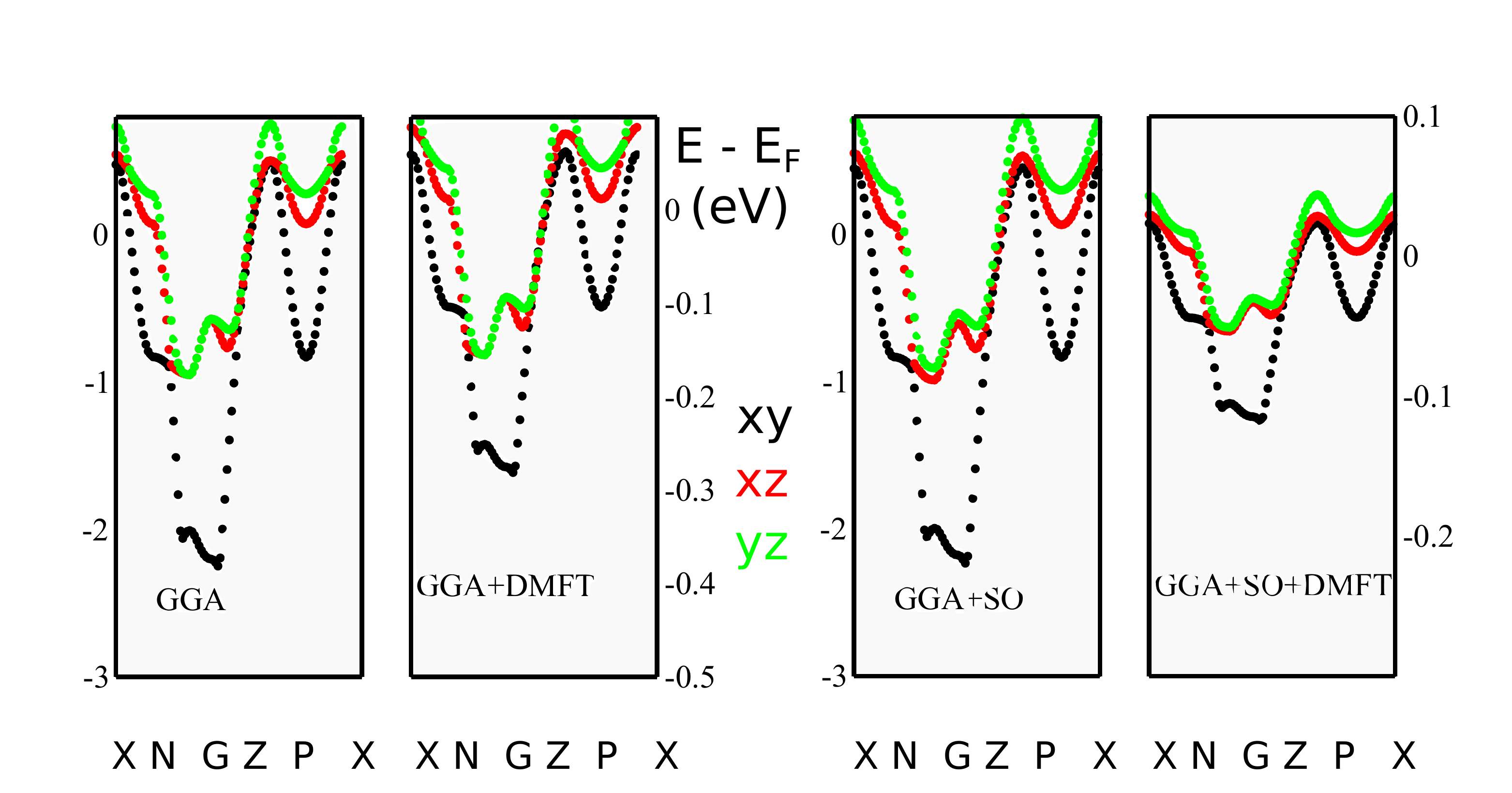}
\caption{DFT(GGA)+SOC results for the multi-orbital band structure of Sr$_{2}$RuO$_{4}$.  Inclusion of multi-orbital correlations leads to a sizable reduction in coherent spectral weight, which is transferred to the incoherent part of the spectral functions (not shown).  From analysis of the corresponding self-energies at $30$~K, we estimate effective mass enhancements of $O(3-4)$ times the band masses for the three $t_{2g}$ orbitals, also in fair accord with dHvA estimates.}
\label{bands}
\end{figure}
 
  The emergent low-energy picture from our normal-state results is the following: already at high $T$,
the Hund coupling results in $S=1$ on each Ru site, which also controls the very
low $T_{FL}\simeq 23$~K via a  combination of Hund metal physics~\cite{georges} and a dimensional crossover (see above).
At high $T$, both the SOC as well as the (small $t_{\perp}~0.1t_{a\sigma,b\sigma'}$) interlayer
hopping are irrelevant, and one can view the system in terms of an orbitally
degenerate quasi-1D (xz-yz) system coupled to a wider 2D xy-band via two-body
Coulomb interactions in the $t_{2g}$-sector.  At low $T<T_{FL}$, both (especially $t_{\perp}$) become
relevant, inducing both, spin-orbital entanglement and a dimensional
crossover (see SI $(ii)$), resulting in an anisotropic quasi-3D correlated FL.
Importantly, since sizable local interactions and SOC have already drastically renormalized the quasi-coherent part of the bare GGA band structure as shown in Fig.~\ref{bands}, it now
follows that the ``effective interactions'' which finally cause cooper pairing should involve heavily renormalized, instead of bare interactions.  Leaving the detailed microscopics for future, we restrict ourselves to qualitative but symmetry-based reasoning.
With the assumption, bolstered by the maximally enhanced interband charge susceptibilities in the $xz,yz$ sector in the normal-state DFT+DMFT calculation, that SC primarily arises from the quasi-1D xz-yz bands,
and noticing that only these orbitals have non-negligible overlap with out-of-plane $t_{2g}$ orbitals
in the BCT structure, we are led to a picture similar to that proposed by Hasegawa {\it et al.}~\cite{hasegawa}.
In the BCT structure, restricting ourselves to symmetry arguments alone leads one to a two-fold degenerate,
multi-band spin-triplet and odd-parity pair state of the form

\be
 \Delta(k)={\bf z}\Delta_{0}(sin\frac{k_{x}a_{0}}{2}cos\frac{k_{y}a_{0}}{2} + icos\frac{k_{x}a_{0}}{2}sin\frac{k_{y}a_{0}}{2}).(cos\frac{k_{z}c_{0}}{2}+a_{1})
\ee
which reduces to the form $\Delta(k)={\bf z}\Delta_{0}(k_{x}+ik_{y})(cos(k_{z}c_{0})+a_{1})$ for
small $k_{x},k_{y}$, where the $a_{1}$ component can exist on symmetry
grounds in Sr$_{2}$RuO$_{4}$~\cite{hasegawa} (indeed, it results from the interband proximity-induced $(k_{x}+ik_{y})$ gap on the $xy$-band).  The form factor
cos$(k_{z}c)$ immediately leads to horizontal line nodes at $k_{z}=\pi/2c\pm \delta k_{0}$ as long as 
$|a_{1}|<1$, as required from various data.  The crucial point we wish to make is that this choice naturally
results from an effective interaction which primarily couples the xz-yz band states
in the BCT structure: the nearly soft interband charge fluctuations can also readily lead to pairing.
This supports the mechanism of Raghu {\it et al.}  The interband proximity effect will
naturally induce a secondary gap of the form $\Delta_{xy}(k)\simeq {\bf z}(k_{x}+ik_{y})$ on the
2D xy-band~\cite{rice}.   However, our results show that the charge susceptibility in the $xy$-band is also comparably enhanced, and this suggests a truly multiband modelling of the SC instability in Sr$_{2}$RuO$_{4}$ involving cooper
 pairs with multiband and mixed-spin character~\cite{simon} is necessary.  
Whether such microscopics lead to a gap function consistent with that above remains to be seen.  However, 
it is interesting to notice that $\Delta(k)$ arrived at from symmetry considerations above describes a 
two-fold degenerate
state with spin-triplet character.  The response to uniaxial strain discovered by Hicks {\it et al.}~\cite{hicks} can then be naturally understood: uniaxial strain will lift this two-fold degeneracy, stabilizing either $p_{x}(p_{y})$ symmetries for tensile(compressive) strain
(this argument cannot rationalize the enhancement of $T_{c}$ under uniaxial strain, which requires full microscopics).
  In fact, strain must also lead to further anisotropies in the normal state responses, since LDA calculations under strain already show a Pomeranchuk-like anisotropy for the xy-($\gamma$)-FS sheet under uniaxial strain.  It would be interesting to probe this aspect in more detail in future.  Moreover, while USC following from quasi-1D xz-yz bands is not of the topological type, the proximity effect restores this aspect~\cite{raghu}.  But given the relative weakness of the proximity coupling, one expects that the edge currents using SQUID imaging studies~\cite{kirtley} will have magnitudes much smaller than expected, precisely as apparently seen.  Thus, a natural choice for an USC arising first in the quasi-1D bands seems to be qualitatively consistent with constraints mandated by a wide range of data.
 However, within the truly multiband view, Scaffidi {\it et al.}~\cite{simon} posit an alternative explanation (high
Chern number) for the much smaller
edge currents in SQUID data, so the issue is unresolved and mandates more study.
In common with most other proposals, this idea still does not fully address $(i)$ the fact that the true situation is
probably somewhere in between the proposals of Rice {\it et al.}~\cite{rice} and Raghu {\it et al.}, and involves all
three $d$-bands (see, however~\cite{simon}), and $(ii)$ more importantly, the requirement imposed by spin-orbital entanglement, namely, that the pair symmetry cannot be classified either as spin singlet or triplet when SOC is strong (the renormalized SOC is in fact larger than the renormalized crystal-field splitting
 in LDA+DMFT studies~\cite{pavarini}.  In fact, very recent spectroscopic evidence~\cite{damascelli1} strongly
 suggests a complicated ${\bf k}$-dependent spin-orbital entanglement.  More evidence obtains from very recent
strain data~\cite{steppke}, where increase of T$_{c}$ under uniaxial strain is suggested to lead to a transition
between two USC states having odd- and even parity: within a scenario of a pair function with an admixture of spin singlet and triplet (due to SO-entanglement), it could be that the relative weight of the singlet-triplet
admixture is modified in favor of an even-parity state under strain.  These requirements demand a generalization of the
argument above to allow all three $d$-bands, and a ${\bf k}$-dependent admixture of spin singlet and triplet
components into the full USC gap function, and is out of scope of this work.
\section*{Conclusion}
  To conclude, we have revisited the important issue of detailed characterization of the ``normal'' state in
Sr$_{2}$RuO$_{4}$ using state-of-the-art first-principles correlation (GGA+SOC+DMFT(CTQMC)) approach.  Armed with
excellent semiquantitative accord with a wide range of normal state transport and magnetic fluctuation data, we
propose that a hitherto unnoticed aspect related to dominant interband charge fluctuations from the $xy,\, xz,\, yz$ bands potentially emerges as a new ``pair glue'' for USC in Sr$_{2}$RuO$_{4}$.  Assuming that these charge fluctuations
 are the major contributors to the pair glue, we use symmetry arguments to argue that USC with a required pair
 symmetry primarily arises on the $\alpha,\beta$-FS sheets, and induces it on the quasi-2D xy-band via an interband
proximity effect.  However, in view of the comparable charge susceptibility from the $xy$-band, its contribution will also be comparable to that of the $xz,\, yz$ bands in reality, necessitating a full three-band scenario in an (orbital-dependent) intermediate-to-strong coupling framework.  Our study establishes a concrete link between normal state responses and the USC instability,
and provides evidence that nearly soft interband charge fluctuations in the Hund metal, involving all Ru $d$ bands potentially play an important role in fomenting USC in this system.

\section*{Acknowledgements}
SA would like to acknowledge UGC (India) for a re-
search fellowship. The authors acknowledge FIST
programme of DST (India) for computational facility at
the Physics Department of IIT Kharagpur.

\end{document}